\documentclass[12pt,preprint]{aastex}
\usepackage{epsfig}
                                                                                
\def\vkm{km s$^{-1}$}

\def\degree{$^\circ$}
\def\arcs#1{$#1''$}
\def\arcsa#1#2{$#1^{\prime\prime}_{^\textrm{.}}#2$}

\def\solarmass{$M_\odot$}

\def\mJyb{mJy beam$^{-1}$}

\def\mJybk{mJy beam$^{-1}$ km s$^{-1}$}

\def\cms{cm$^{-2}$}

\def\VLSR{V_\textrm{\scriptsize LSR}}
\def\Vsys{V_\textrm{\scriptsize sys}}
\def\Voff{V_\textrm{\scriptsize off}}

\def\mH2{m_{\textrm{\scriptsize H}_2}}

\def\H2{H$_2$}
\def\N2HP{N$_2$H$^+$}
\def\HCOP{HCO$^+$}

\def\NH3{NH$_3$}

\def\DtCO{D$_2$CO}
\def\CHtDOH{CH$_2$DOH}
\def\CHtOH{CH$_3$OH}
\def\NHtCHO{NH$_2$CHO}
\def\CHtSH{CH$_3$SH}

\def\SOt{$N_J=8_9-7_8$}

\def\HCOP{HCO$^+$}

\def\putfig#1#2#3{\epsfig{scale=#1,angle=#2,figure=#3}}
\def\putfiga#1#2#3{}
\def\leftblank#1{}
\def\bf#1{}

\begin{document}

\title{Formation and Atmosphere of Complex Organic Molecules of the HH 212
Protostellar Disk}

\author{Chin-Fei Lee\altaffilmark{1,2}, Zhi-Yun Li\altaffilmark{3}, Paul
T.P.  Ho\altaffilmark{1,4}, Naomi Hirano\altaffilmark{1}, Qizhou
Zhang\altaffilmark{4}, and Hsien Shang\altaffilmark{1}}


\altaffiltext{1}{Academia Sinica Institute of Astronomy and Astrophysics,
P.O. Box 23-141, Taipei 106, Taiwan; cflee@asiaa.sinica.edu.tw}
\altaffiltext{2}{Graduate Institute of Astronomy and Astrophysics, National Taiwan
   University, No.  1, Sec.  4, Roosevelt Road, Taipei 10617, Taiwan}
\altaffiltext{3}{Astronomy Department, University of Virginia, Charlottesville, VA 22904}
\altaffiltext{4}{Harvard-Smithsonian Center for Astrophysics, 60 Garden
Street, Cambridge, MA 02138}


\begin{abstract}

HH 212 is a nearby (400 pc) Class 0 protostellar system recently found to
host a ``hamburger"-shaped dusty disk with a radius of $\sim$ 60 AU, deeply
embedded in an infalling-rotating flattened envelope.   We have
spatially resolved this envelope-disk system with the Atacama Large
Millimeter/submillimeter Array at up to $\sim$ 16 AU (\arcsa{0}{04})
resolution.  The envelope is detected in \HCOP{} J=4-3 down to the dusty
disk.  Complex organic molecules (COMs) and doubly deuterated formaldehyde
(\DtCO{}) are detected above and below the dusty disk within $\sim$ 40 AU of
the central protostar.  The COMs are methanol (\CHtOH{}), deuterated
methanol (\CHtDOH{}), methyl mercaptan (\CHtSH{}), and formamide (\NHtCHO{},
a prebiotic precursor).  We have modeled the gas kinematics in \HCOP{} and
COMs, and found a centrifugal barrier at a radius of $\sim$ 44 AU, within
which a Keplerian rotating disk is formed.  This indicates that \HCOP{}
traces the infalling-rotating envelope down to centrifugal barrier and COMs
trace the atmosphere of a Keplerian rotating disk within the centrifugal
barrier.  The COMs are spatially resolved for the first time, both radially
and vertically, in the atmosphere of a disk in the earliest, Class 0 phase
of star formation. Our spatially resolved observations of COMs favor their
formation in the disk rather than a rapidly infalling (warm) inner envelope. 
The abundances and spatial distributions of the COMs provide strong
constraints on models of their formation and transport in low-mass star
formation.

\end{abstract}


\section{Introduction}

In theory of star formation, rotationally supported disks are expected to
form inside collapsing cores, feeding the protostars at the center.  They
will have a Keplerian rotation if the mass of the protostars dominates that
of the disks.  In models of non-magnetized core collapse, a Keplerian disk
can indeed form as early as in the Class 0 phase \citep{Terebey1984}. 
However, a realistic model should include magnetic field, because recent
survey toward a few Class 0 sources shows that molecular cores are
magnetized and likely to have an hourglass B-field morphology
\citep{Chapman2013}.  Unfortunately, in many current models of magnetized
core collapse, the magnetic field produces an efficient magnetic braking
that removes the angular momentum and thus prevents a Keplerian disk from
forming at the center \citep{Allen2003,Mellon2008}.  In those cases, only a
flattened envelope called the pseudodisk can be formed around the central
source \cite[e.g.,][]{Allen2003}.  Magnetic-field-rotation misalignment is
sometimes able to solve this so-called magnetic braking catastrophe
\citep{Joos2012,Li2013}.

Before the advent of the Atacama Large Millimeter/submillimeter Array
(ALMA), a few Class 0 disk candidates in, e.g., HH 212
\citep{Codella2007,Lee2008}, HH 211 \citep{Lee2009}, L1527
\citep{Tobin2012}, and VLA 1623 \citep{Murillo2013a}, have been identified. 
With ALMA, we have started to resolve their kinematics, which are found to
be roughly Keplerian \citep{Murillo2013,Sakai2014}.  In addition, we can
also map the surrounding envelopes and study the transition from the
envelopes to the disks \citep{Sakai2014,Lee2014}.  Recent study at high
resolution has even started to resolve the transition region between the
envelope and disk in L1527 \citep{Sakai2017}.

HH 212 is a well-studied Class 0 protostellar system with evidence for a
Keplerian disk resolvable with ALMA, and thus a good target to study the
disk formation and the transition from the envelope to the disk.  It is
deeply embedded in a compact molecular cloud core in the L1630 cloud of
Orion, which is at a distance of about 400 pc \citep{Kounkel2017}. 
The central source is the Class 0 protostar IRAS 05413-0104, with a
bolometric luminosity $L_\textrm{\scriptsize bol}\sim$ 9 $L_\odot$ (updated
for the distance of 400 pc) \citep{Zinnecker1992}.  It drives a powerful
bipolar jet \citep{Zinnecker1998,Lee2007}.  A Keplerian disk must have
formed in order to launch the jet, according to current models of jet
launching.

In previous ALMA Cycle 0 observations at $\sim$ 200 AU (\arcsa{0}{5})
resolution, we have detected a flattened envelope in \HCOP{} with an outer
radius of $\sim$ 1000 AU around the central protostar \citep{Lee2014}.  The
envelope is infalling toward the center, spinning up at smaller radii in a
manner that is consistent with angular momentum conservation.  It can thus
be identified as a pseudodisk in current models of the magnetized core
collapse \citep{Allen2003,Mellon2008}.  Based on the infall and rotation
velocity profiles, we estimated a centrifugal radius of $\sim$ 120 AU
(\arcsa{0}{3}), thus a Keplerian disk is expected to have formed closer in,
with a radius $<$ 120 AU.  A similar disk radius was estimated based on the
gas kinematics in C$^{17}$O toward the center \citep{Codella2014}.  Recently
with ALMA observations at a resolution of $\sim$ 8 AU (\arcsa{0}{02}), we
have not only detected the disk but also spatially resolved the disk
structure in dust continuum \citep{Lee2017}.  The dusty disk has a radius of
$\sim$ 60 au, which is half of the centrifugal radius estimated before at
low resolution.  A natural question is: Is this the radius of the
centrifugal barrier as found in L1527 \citep{Sakai2014}?  In addition, the
disk structure is resolved in vertical direction, showing a flared structure
within $\sim$ 40 au of the central protostar.  Could this radius be the
radius of the Keplerian disk?  Without detailed kinematic study of the
envelope and disk, we can not determine the radius at which the Keplerian
disk has formed.


In this paper, we will present our ALMA observations in \HCOP{} J=4-3 at
$\sim$ 24-40 AU resolutions, which is $\sim$ 5-8 times higher than the
previous observations, in order to resolve the envelope structure and
kinematics.  At these high resolutions, we can resolve and model the infall
velocity profile directly.  Moreover, we also detect several complex organic
molecules (COMs) in the disk atmosphere at $\sim$ 16 AU resolutions,
including methanol (\CHtOH), deuterated methanol (\CHtDOH), methyl mercaptan
(\CHtSH), and formamide (\NHtCHO).  COMs have been detected before around
the centrifugal barrier or disks/envelopes in other protostellar systems
\citep{Sakai2014,Oya2016,Jorgensen2016}.  Here, we not only detect them but
also spatially resolve their structure.  Our observations not only provide
conclusive evidence that our detected COMs are associated with the disk (as
opposed to the rapidly infalling envelope or large-scale outflow), at least
for this particular Class 0 source, but also enable us to show that the disk
is rotating, with a profile consistent with Keplerian.  The disk rotation
would be difficult to determine without the COMs in the disk atmosphere
above and below the optically thick dusty disk.  Together, the new
observations enable us to better discuss the transition of the envelope to
the disk, and the formation and kinematics of the disk.

\section{Observations}\label{sec:obs}

Observations of the HH 212 protostellar system were carried out with ALMA in
Band 7 at $\sim$ 350 GHz in Cycles 1 and 3, with 32-45 antennas (see Table
\ref{tab:obs}).  The Cycle 1 project was carried out with 2 executions, both
on 2015 August 29 during the Early Science Cycle 1 phase.  The projected
baselines are 15-1466 m.  The maximum recoverable size scale is $\sim$
\arcsa{2}{5}.  A 5-pointing mosaic was used to map the system within $\sim$
\arcs{15} of the central source at an angular resolution of $\sim$
\arcsa{0}{16} (64 AU).  The Cycle 3 project was carried out with 2
executions in 2015, one on November 5 and the other on December 3, during
the Early Science Cycle 3 phase.  The projected baselines are 17-16196 m. 
The maximum recoverable size scale is $\sim$ \arcsa{0}{4}.  One pointing was
used to map the center of the system at an angular resolution of $\sim$
\arcsa{0}{02} (8 AU).  For the Cycle 1 project, the correlator was set up to
have 4 spectral windows, with one for CO $J=3-2$ at 345.795991 GHz, one for
SiO $J=8-7$ at 347.330631 GHz, one for HCO$^+$ $J=4-3$ at 356.734288 GHz,
and one for the continuum at 358 GHz with possible other weak lines (see
Table \ref{tab:corr1}).  For the Cycle 3 project, the correlator was more
flexible and thus was set up to include 2 more spectral windows, with one
for SO \SOt{} at 346.528481 GHz and one for H$^{13}$CO$^+$ $J=4-3$ at
346.998338 GHz (see Table \ref{tab:corr3}).  The total time on the HH 212
system is $\sim$ 148 minutes.

The data were calibrated with the CASA package for the passband, flux, and
gain (see Table \ref{tab:calib}).  In this paper, we only present the
observational results in \HCOP{}, \DtCO, and 4 COMs (see Table
\ref{tab:lines}),  with \HCOP{} tracing the envelope, \DtCO{}
and COMs tracing the disk around the central source.  The
velocity resolution is $\sim$ 0.212 \vkm{} per channel for the lines in the
spectral line windows and $\sim$ 0.848 \vkm{} per channel for the lines in
the continuum window.  Cycle 1 data and Cycle 3 data are combined to make
the maps.  For \HCOP{} maps, the data with the $uv$-distance greater than
3000 m (with a corresponding angular scale of $\sim$
\arcsa{0}{06}) are excluded because no \HCOP{} emission is detected there. 
We used a robust factor of 0.5 for the visibility weighting to generate the
\HCOP{} maps at $\sim$ \arcsa{0}{06} resolution.  In addition, we convolved
the maps to \arcsa{0}{1} resolution to map the emission at low velocity,
which is more extended.  The noise levels can be measured from line-free
channels and are found to be $\sim$ 3.1 \mJyb{} (or $\sim$ 10 K) for $\sim$
\arcsa{0}{06} resolution and 3.9 \mJyb{} (or $\sim$ 3.75 K) for $\sim$
\arcsa{0}{1} resolution.  For the maps of the complex organic molecules, we
used a robust factor of 0.5 and the data with the $uv$-distance shorter than
8000 m (with a corresponding angular scale of $\sim$
\arcsa{0}{02}), because of no detection at longer $uv$-distance.  The
resulting resolutions are $\sim$ \arcsa{0}{04}.  The noise levels are found
to be $\sim$ 1.7 \mJyb{} (or $\sim$ 11 K) for 0.848 \vkm{} and 3.0 \mJyb{}
(or $\sim$ 20 K) for 0.212 \vkm{}.  The velocities in the channel maps are
LSR velocities.

\section{Results}

The systemic velocity in HH 212 is assumed to be $\Vsys= 1.7\pm0.1$ \vkm{}
LSR, as in \citet{Lee2007}.  Throughout this paper, in order to facilitate
our presentations, we define an offset velocity $\Voff = \VLSR - \Vsys$ and
rotate our maps by 23\degree{} clockwise to align the jet axis in the
north-south direction.  The jet has an inclination of $\sim$ 4\degree{}
to the plane of the sky, with the northern component tilted toward us
\citep{Claussen1998}.  A dusty disk has been detected perpendicular to the jet axis
and found to be nearly
edge-on with the nearside titled slightly by $\sim$ 4\degree{} to the south
\citep{Lee2017}.

\subsection{Flattened Envelope in \HCOP{}} \label{sec:env}

The envelope can now be better studied at higher angular resolutions.  As
before, \HCOP{} emission is detected near the equatorial plane within
$\sim$ 3 \vkm{} of the systemic velocity.   Based on our analysis of the
gas kinematics below and later in Section \ref{sec:mod}, the emission there
shows an infall motion with some rotation and thus traces a flattened
infalling-rotating envelope. In order to see the velocity distribution
pictorially, we divide the velocity range into a low velocity range with
$|\Voff| \leq 1.5$ \vkm{} and a high velocity range with $1.5\leq |\Voff|
\leq 3$ \vkm{}.

\begin{figure} [!hbp]
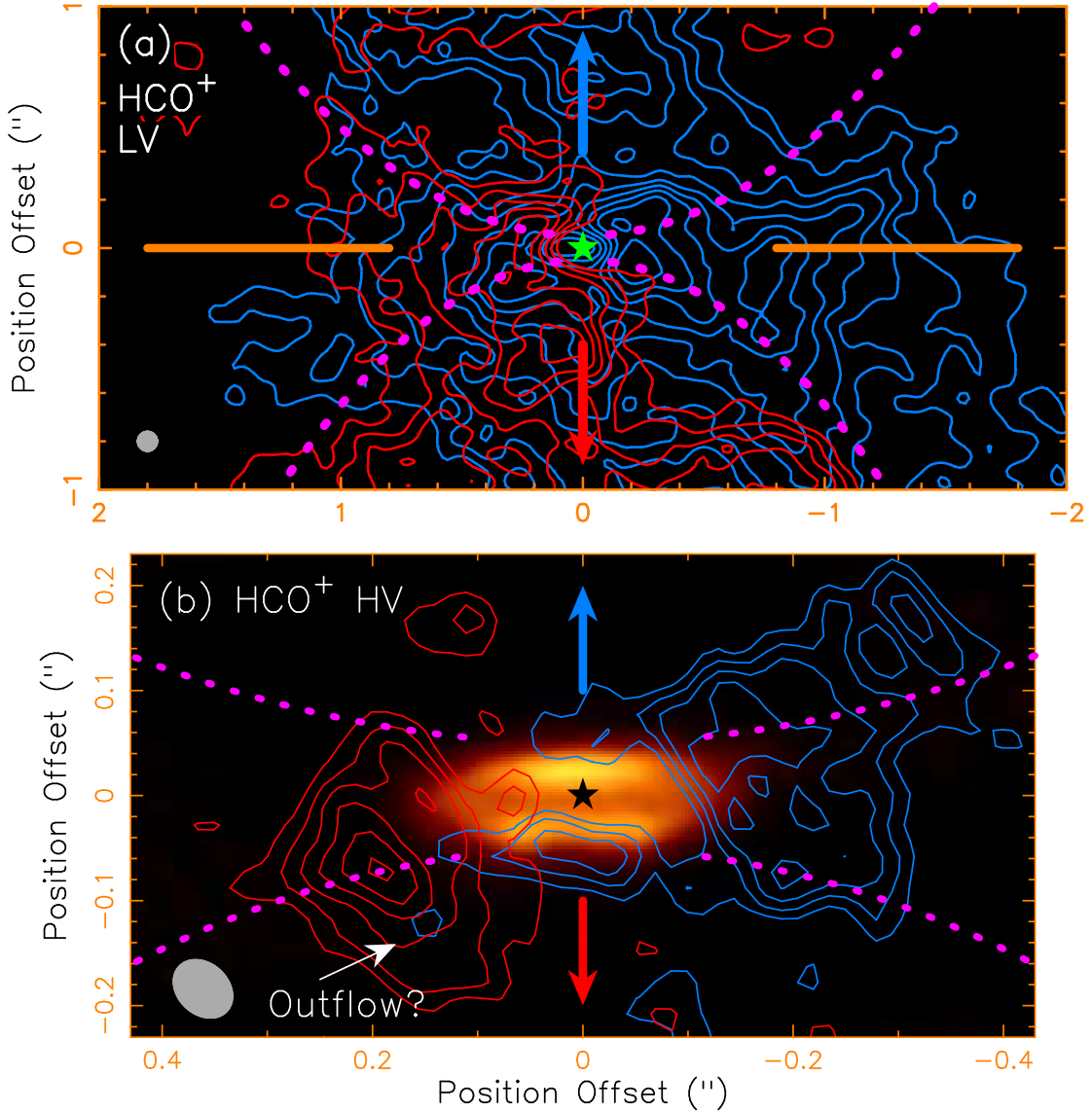

\centering
\putfig{0.9}{270}{f1.eps} 
\figcaption[]
{ALMA \HCOP{} maps toward the center of the HH 212 system, rotated by
22.5\degree{} clockwise to align the jet axis in the north-south direction.
The star marks the position of the central protostar.
The red and blue arrows indicate the axes of the redshifted component
and blueshifted component of the jet, respectively. 
The jet has an inclination angle of $\sim$ 4\degree{} to the plane of the sky.  
The dotted parabolic curves delineate roughly the upper and lower boundaries of
the \HCOP{} envelope, as defined in Section \ref{sec:env}.
(a) shows the maps at low redshifted velocity 
($\Voff$ $\sim$ 0 to 1.5 \vkm{}) and low blueshifted velocity 
($\Voff$ $\sim$ -1.5 to 0 \vkm{}) at \arcsa{0}{1} resolution. 
The contour levels start at 4$\sigma$ with a step of 2$\sigma$, where $\sigma=5$ \mJybk{}.
The orange lines mark the equatorial plane.
 (b) shows the maps at high redshifted velocity ($\Voff$ $\sim$ 1.5 to 3.0 \vkm{})
and high blueshifted velocity ($\Voff$ $\sim$ -3.0 to -1.5 \vkm{}) 
at \arcsa{0}{06}$\times$\arcsa{0}{05}
resolution, on top of the 351 GHz continuum map of the dusty disk adopted
from \citet{Lee2017}.  
The contour levels start at 3$\sigma$ with a step of 1$\sigma$, where $\sigma=3$ \mJybk{}.
\label{fig:HCOP}}
\end{figure}

Figure \ref{fig:HCOP}(a) shows the \HCOP{} maps at low velocity at
\arcsa{0}{1} resolution.  The redshifted and blueshifted emissions near the
equatorial plane (indicated by the orange lines) within a radius of $\sim$
\arcs{2} of the central source trace the flattened envelope.  Figure
\ref{fig:HCOP}(b) shows the \HCOP{} maps at high velocity at \arcsa{0}{06}
resolution.  At high velocity, since the emission structure is more compact,
a higher resolution is used.  The innermost part of the envelope can be seen
at high velocity down to $\sim$ \arcsa{0}{1} of the central source.  As can
be seen from these two figures, the envelope is flared with a thickness
increasing with increasing distance from the central source.  For
simplicity, the boundaries of the envelope are assumed to have a parabolic
structure, with $z=a+bR^2$ in the cylindrical coordinate system.  Here $a$
is the position offset from the central source and $b$ is a constant
describing the curvature.  Using a rough eye fitting to the redshifted and
blueshifted \HCOP{} emission of the envelope at both low velocity (Figures
\ref{fig:HCOP}a) and high velocity (Figure \ref{fig:HCOP}b), we obtained
$a=0.05$ and $b=0.45$ for the upper boundary and $a=-0.05$ and $b=-0.6$ for
the lower boundary, as delineated by the dotted parabolic curves.  Note that
since the boundaries are neither symmetric nor sharply defined, the
parabolic curves are mainly to guide the readers.

At low velocity as shown in Figure \ref{fig:HCOP}(a), the blueshifted
emission of the envelope is seen across the central source, with the west
side brighter than the east side.  The redshifted emission is weaker and is
seen mainly in the east.  As discussed in \cite{Lee2014}, this spatial
distribution indicates that the envelope has mainly infall motion but some
small rotation with the redshifted side in the east and blueshifted side in
the west.  The redshifted emission is fainter and even absent in the west,
due to a self-absorption in the near side of the infalling envelope
\citep{Evans1999}.  An emission hole is seen at the center toward the dusty
disk  in both blueshifted and redshifted maps.  This is because the
dusty disk is bright and optically thick \citep{Lee2017}, thus the emission
behind it is blocked and the emission in front of it appears absorbed
against the bright background.  The emission above the upper envelope
boundary and below the lower envelope boundary traces the outflow and
jet-like emission, and will be discussed in a future publication.



At high velocity as shown in Figure \ref{fig:HCOP}b, the redshifted and
blueshifted emissions are now seen on the opposite sides within $\sim$
\arcsa{0}{3} of the central source surrounding the dusty disk, indicating
that the motion there becomes dominated by the rotation.  No emission is
detected toward the dusty disk within $\sim$ \arcsa{0}{1} of the center,
because the disk continuum emission becomes optically thick there
\citep{Lee2017}.  The redshifted emission extending slightly to the south from the
lower boundary could trace a low-velocity outflow coming out of the
envelope.  The blueshifted emission above the upper disk surface and below
the lower disk surface is from the infalling envelope in the farside that is
not blocked by the disk.



\begin{figure} [!hbp]
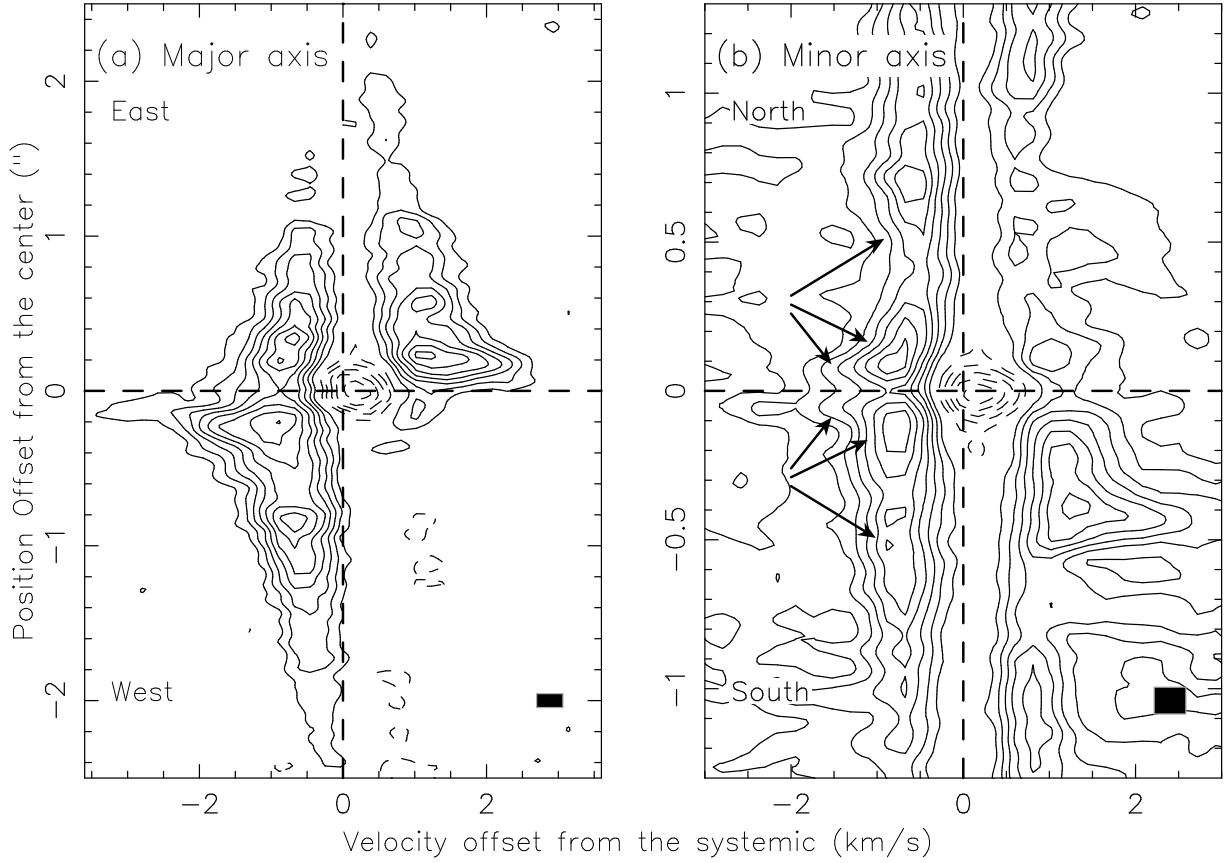

\centering
\putfig{0.65}{270}{f2.eps} 
\figcaption[]
{PV diagrams of the \HCOP{} envelope, along (a) the major axis and (b) the
minor axis.  These diagrams are used to study the rotation and infall
motion in the envelope.  The horizontal dashed line indicates the position
of the protostar.  The vertical dashed line indicates the systemic velocity. 
The arrows in (b) point to the PV structure of the envelope.  The contour
levels start at 3$\sigma$ with a step of 1$\sigma$, with $\sigma=3.7$ K. 
The boxes in the lower left corners show the velocity and angular
resolutions of the PV diagrams.
\label{fig:pvHCOP}}
\end{figure}

Figure \ref{fig:pvHCOP}a shows the position-velocity (PV) diagram of the
envelope cut along the major axis (equator).  It shows a blueshifted
triangular structure pointing toward the high blueshifted velocity and a
redshifted triangular structure pointing toward the high redshifted
velocity, similar to those seen before at a lower resolution of $\sim$
\arcsa{0}{45} \citep{Lee2014}.  As discussed by the authors, these features
are signatures of infall motion in the envelope, with the blueshifted
triangular structure coming from the farside of the envelope and the
redshifted from the nearside.  The blueshifted triangular structure shifts
slightly to the west because of a small rotation (going from the east to the
west in the farside) in the infalling envelope.  The base of the redshifted
triangular structure near the systemic velocity is absent due to the
self-absorption in the nearside, which is infalling toward the central
source and thus preferentially absorbs the redshifted part of the emission
(Evans 1999).  This absorption by infalling envelope on the nearside is the
reason why there is little redshifted HCO+ emission on the west side of the
central source.  At higher angular resolution of $\sim$ \arcsa{0}{1}, the
high redshifted ($\gtrsim 1.5$ \vkm{}) and high blueshifted ($\lesssim -2$
\vkm) emissions near the central source are now seen on the opposite sides
of the source, confirming that the motion there becomes dominated by the
rotation.

At high resolution of $\sim$ \arcsa{0}{1}, we can study the infall
motion directly using the PV diagram of the envelope cut along the minor
axis, as shown in Figure \ref{fig:pvHCOP}b.  The PV structure for the faint
emission at high velocity with $|\Voff| > 2$ \vkm{} should be mainly from
the outflow and jetlike emission along the jet axis, as seen in Figure 7b in
\citet{Lee2014}.  In addition, the emission on the redshifted side suffers
significantly from the self-absorption.  Therefore, we only focus on the low
velocity part ($|\Voff| < 2$ \vkm{}) on the blueshifted side as pointed by
the arrows, and model it in Section \ref{sec:mod} below to see if the
infall velocity increases toward the center.

\subsection{Tracing Disk Atmosphere with COMs}

\begin{figure} [!hbp]
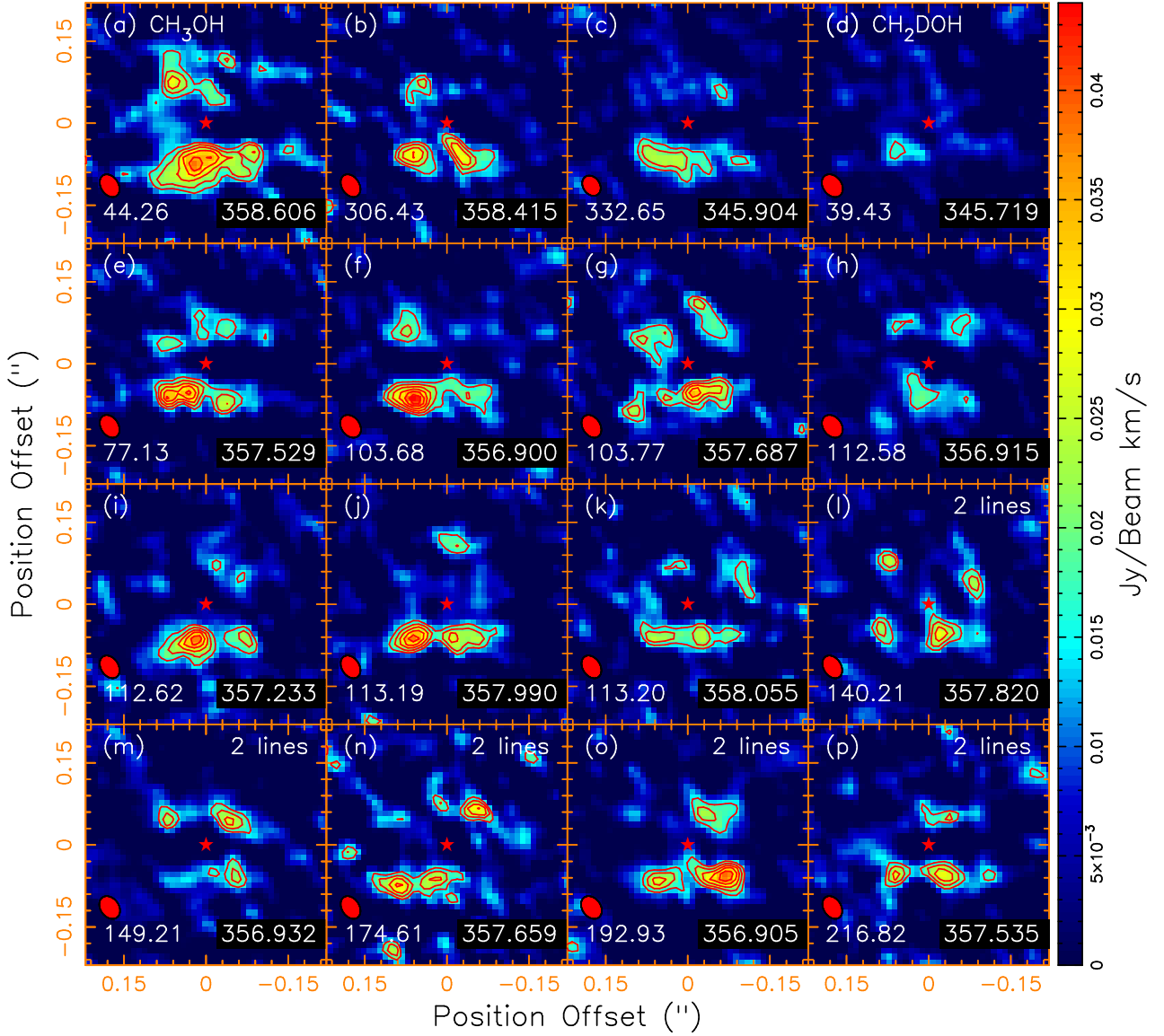

\centering
\putfig{0.9}{270}{f3.eps} 
\figcaption[]
{Integrated intensity maps of methanol ((a)-(c)) and deuterated methanol ((d)-(p))
at various line transitions in the order of increasing upper energy level, 
integrated from $\Voff$ of $-$4 to 4 \vkm{}.
Upper energy level (K) and rest frequency (GHz) of each transition
are indicated in the lower left and right corners, respectively.
Panels (l)-(p) each contain 2 hyperfine lines. The star marks
the protostar position. The noise levels are $\sim$ 5-7 \mJybk{}.
The contours start from 15 \mJybk{} with a step of 5 \mJybk{}.
\label{fig:methanol}}
\end{figure}

Line emissions from 4 complex organic molecules (COMs) and \DtCO{} are
detected toward the center (see Table \ref{tab:lines}).  Figure
\ref{fig:methanol} shows the integrated intensity maps of \CHtOH{}
(methanol, 3 lines) and \CHtDOH{} (deuterated methanol, 18 lines at 13
frequencies) lines in various transitions.  Since the emission morphology is
similar, the maps of the same molecule are stacked together to make a map
with a higher S/N ratio.  Figure \ref{fig:complexmol} shows the stacked maps
of \CHtOH{} and \CHtDOH{}, along with the maps of 3 other molecules,
\CHtSH{} (methyl mercaptan), \NHtCHO{} (formamide), and \DtCO{} (doubly
deuterated formaldehyde), all at $\sim$ \arcsa{0}{04} resolution.






Unlike \HCOP{}, the molecules \CHtOH{}, \CHtDOH{}, and \CHtSH{} are only
detected within $\sim$ 40 AU (\arcsa{0}{1}) of the central source, and the
molecules \NHtCHO{} and \DtCO{} are only detected closer in (within $\sim$
30 AU).  Notice that \CHtDOH{} is a D-substituted methanol and \CHtSH{} is a
sulfur analog of methanol, and they both have the same spatial distribution
as the methanol.  The emission structures are more extended for
\CHtOH{}, \CHtDOH{}, and \CHtSH{} than for \NHtCHO{} and \DtCO{}.  The
A-coefficients of these 5 molecules can be roughly divided into two groups,
one for \CHtOH{}, \CHtDOH{}, and \CHtSH{} having a value of $\sim$ 10$^{-4}$
s$^{-1}$, and the other for \NHtCHO{} and \DtCO{} having a value of
(1$-$3)$\times10^{-3}$ s$^{-1}$, as seen in Table \ref{tab:lines}.  Thus,
the difference in the spatial distribution of these molecules might be
partly due to the different values of their A-coefficients
{\bf which could reflect different critical densities.} These complex
molecules are detected at $\sim$ \arcsa{0}{05} (20 AU) above and below the
midplane, and thus above the upper dusty disk surface in the north and below
the lower dusty disk surface in the south.  The observed morphologies are
suggestive of the COMs tracing the outer layers (i.e., atmosphere) of the
disk, although it is also plausible that they trace the base of a disk wind. 
We favor the former interpretation over the latter because, as we will show
below, the radial component of the velocity for these layers is much smaller
than the rotational component.   In addition, based on our analysis of
the gas kinematics below and later in Section 3.3, the emission there shows
a Keplerian rotation and thus indeed traces the atmosphere of a Keplerian
rotating disk. Again, no molecular line emission is detected toward the
opaque dusty disk.  Since the dusty disk is optically thick, we can not
determine whether this is because of an absorption against the bright dust
continuum or because of no emission of these molecules in the disk.  Since
the disk midplane has a temperature of $\sim$ 70 K at $\sim$ 40 AU
\citep{Lee2017}, these molecules could also be depleted onto the dust grains
near the midplane at the observed radius.  The emission is brighter in the
south than in the north.  This is in opposite to the dust emission, which is
brighter in the north.  The redshifted emission and blueshifted emission are
seen on the opposite sides of the jet axis, confirming that the disk
atmosphere traced by the COMs is rotating.  Notice that in the south in the
methanol and deuterated methanol maps, the blue and red emissions have some
overlap at small radii, due to the low velocity resolution and (thermal and
possibly turbulent) line broadening, as discussed later.  The emission maps
show mainly a two-peak structure in the atmosphere, with one peak to the
east and the other to the west of the rotational axis, tracing the two
limb-brightened edges of a rotating ring there, as discussed later. 
\NHtCHO{} and probably \DtCO{} could trace a rotating ring closer to the
rotational axis than \CHtOH{}, \CHtDOH, and \CHtSH{}, because their emission
peaks are closer in.

The stacked \CHtOH{} and \CHtDOH{} maps have sufficient S/N ratio for
kinematic study.  Figures \ref{fig:pvCH2DOH}a and \ref{fig:pvCH2DOH}b show
the PV diagrams obtained by cutting their emission across the upper (blue)
and lower (red) disk atmospheres parallel to the disk major axis.  Here the
upper disk atmosphere means the atmosphere in the north above the dusty disk
surface, and the lower disk atmosphere means the atmosphere in the south
below the dusty disk surface.  The PV structure is similar in these two
molecules, in agreement with them coming from the same region in the disk
atmosphere.  In addition, the PV structures in the upper and lower disk
atmospheres are also similar, indicating that the atmospheres on both sides
of the disk have a similar motion.  At low velocity ($|\Voff| \lesssim 2$
\vkm{}), linear PV structures are seen across the rotational (jet) axis, as
indicated by the solid lines, which are obtained from a single linear
fit to the linear PV structures seen in both methanol and deuterated
methanol. At the two ends of the linear structures, the velocity becomes
differential and increases toward the center.  We can map these PV
structures using the stacked deuterated methanol map, which has a higher S/N
ratio than the stacked methanol map.  Figure \ref{fig:com}a shows the maps
for both the linear part (green contours) and differential part (red and
blue for redshifted and blueshifted emission, respectively).  For the linear
part, the map shows a band of emission above the upper dusty disk surface
and a band below the lower dusty disk surface.  For the differential PV
part, both the blue and redshifted emission are seen on the two edges of the
disk atmosphere.  In the south, the emissions also extend inward to the
rotational axis.  As shown by the magenta lines, their scale height
decreases as they go closer to the rotational axis, indicating that the disk
atmosphere traced by COMs is flared, as discussed below.  In addition,
their thickness (FWHM) also decreases toward the rotational axis.

Figure \ref{fig:com}b shows a cartoon to explain the PV and the emission
structures of the line emission, by adding a disk atmosphere (outlined by
the curves and lines) on the dusty disk (color disk, with the color changing
from blue to orange for the cold to hot disk surface).  As mentioned
earlier, the disk is almost edge-on with its nearside tilted slightly by
4\degree{} to the south.  The disk atmosphere is assumed to be flared with
the scale height (magenta lines) increasing with the radius, like the dusty
disk.  As can be seen from the cartoon, we can only detect the outermost
ring of the disk atmosphere, which is not blocked by the dusty disk.  For
the upper disk atmosphere, the front part of the ring is projected onto the
dusty disk due to the tilt, and thus is absorbed against the bright disk
continuum emission.  Hence, the observed linear PV structure and its
associated emission are from the back part of the ring.  For the lower disk
atmosphere, the back part of the ring is blocked by the disk, and thus the
observed linear PV structure and its associated emission are from the front
part of the ring.  The radial (infall or expansion) motion of the ring can
be studied with the PV diagrams of methanol and deuterated methanol along
the minor axis, as shown in Figures \ref{fig:pvCH2DOH}c and
\ref{fig:pvCH2DOH}d.  It it clear from the PV diagrams that the infall
velocity there has dropped to zero.  It has a small range of velocities,
probably due to thermal and turbulent (due to, e.g., shock) broadening in
the line emission and the low velocity resolution of $\sim$ 0.85 \vkm{} in
the stacked maps.

Besides the outermost ring, the part of the disk atmosphere along the major
axis (magenta lines) can also be detected, as shown in Figure
\ref{fig:com}b.  This part produces the blue and redshifted emission in the
disk atmosphere extending inward to the rotational axis, with a scale height
decreasing toward the center, associated with the differential PV structure. 
For the upper disk atmosphere, the inner part is not detected because of the
absorption against the dusty disk.  For the lower disk atmosphere, the inner
part can be detected, because the dusty disk, with density decreasing with
height \citep{Lee2017}, becomes optically thin at the highest
rim.  This allows us to see the atmosphere further in, as seen in the
observations.




\subsection{Modeling the Kinematics of the Envelope and Disk} \label{sec:mod}

In this section, we adopt the toy model developed by \citet{Sakai2014},
which has been used to reproduce the similar PV structures of the envelope
and disk in L1527.  In this model, the gas motion in the envelope is
approximated by the motion of a particle with conservation of both specific angular
momentum and total energy. 
If the total energy is negligible at large distances and the specific angular
momentum $l$ is conserved in the infalling material, then the radial velocity
$v_r$ and the rotation velocity $v_\phi$ (where $\phi$ is the azimuthal
angle) at radius $r$ from the protostar are given by:
\begin{equation}
v_r=-\sqrt{\frac{2GM}{r}-\frac{l^2}{r^2}}
\end{equation}
and
\begin{equation}
v_\phi=\frac{l}{r}
\end{equation}
where $M$ is the mass of the central protostar plus disk.
The radius at which the gravitational acceleration is balanced by the
centrifugal force is given by
\begin{equation}
r_c=\frac{l^2}{GM}
\end{equation}
This is the centrifugal radius, where the radial velocity is equal to the rotation velocity,
both with a magnitude of $GM/l$.
The radius at which all the kinetic energy is converted to the rotational motion
is given by
\begin{equation}
r_0=\frac{l^2}{2GM}
\label{eq:CB}
\end{equation}
This is the radius of the centrifugal barrier, where $v_\phi=2GM/l$. It is
half of the centrifugal radius.  The infalling and rotating gas cannot move
inward of the centrifugal barrier, unless it loses kinetic energy and
angular momentum.  If the gas does lose kinetic energy and angular momentum,
then a disk is expected to form with a Keplerian rotation.



The envelope outside the disk is flared with the structure outlined by the
dotted parabolic curves in Figure \ref{fig:HCOP}a.  For given $l$ and $M$,
we can derive model PV structures for the envelope along the major and minor
axis, by calculating the highest and lowest velocities at each position
offset from the central source.  These model PV structures can then be used
to fit the outer boundaries of the observed PV structures.  From
$M$, we also can calculate the Keplerian rotation for the disk.  Thus, the
observed kinematics of the envelope and disk atmosphere can be fitted
simultaneously with only two parameters, $l$ and $M$.

The best-fit parameters to both the envelope and disk kinematics
simultaneously are found to be $l=140\pm30$ AU \vkm{} and $M=0.25\pm0.05$
\solarmass{}, by matching (1) the outer boundaries of the PV structures of
the envelope along the major and minor axis, (2) the maximum velocity in the
PV diagram of the envelope along the major axis at the centrifugal barrier,
(3) the maximum infall velocity of the envelope at the protostar position,
and (4) the differential PV structure seen in the disk.  For the envelope,
the fit is based more on the blueshifted side, because the redshifted side
suffers from self-absorption.  As can be seen in Figures \ref{fig:pvmodel}a
and \ref{fig:pvmodel}b, the model PV structures delineate the outer
boundaries of the observed PV structures reasonably well on the blueshifted
side, and also appear reasonable on the redshifted side.  The best-fit
parameters result in a centrifugal barrier (CB) at $\sim$
\arcsa{0}{11}$\pm$\arcsa{0}{02} with a maximum velocity of $\sim$ 3.2
\vkm{}, roughly matching that seen in the observed PV structure in Figure
\ref{fig:pvmodel}a.  The maximum infall velocity is $\sim$ 1.6 \vkm{},
similar to that seen at the source position in Figure \ref{fig:pvmodel}a and
similar to the maximum velocity in Figure \ref{fig:pvmodel}b.  For the disk,
as marked in the figure, the centrifugal barrier (CB) roughly matches the
two end positions of the linear PV structures seen in methanol and
deuterated methanol, and the Keplerian rotation also roughly matches the
differential PV structures on both the redshifted and blueshifted sides.  In
summary, this simple toy model can broadly reproduce all the important
features in the PV diagrams of the envelope and disk.  The observed
infall velocity indeed increases toward the center, as expected for a
gravitational collapse.


\subsection{Physical condition in the disk atmosphere}

With 18 deuterated methanol lines (at 13 frequencies) detected, we can
construct a population diagram to estimate the mean excitation temperature
and the column density of the molecule in the region traced by the COMs.  It
is a diagram that plots the column density per statistical weight in the
upper energy state in the optically thin limit, $N_u^\textrm{\scriptsize
thin}/g_u$, versus the upper energy level $E_u$ of the lines.  Here
$N_u^\textrm{\scriptsize thin}=(8\pi k\nu^2/hc^3 A_{ul}) W$, where the
integrated line intensity $W=\int T_B dv $ with $T_B$ being the brightness
temperature.


The integrated line intensity of each transition can be measured toward the
lower disk atmosphere where the emission is better detected, using the total
intensity map (integrated over velocity, as shown in Figure
\ref{fig:methanol}) with a 2$\sigma$ cutoff.  The resulting population
diagram is shown in Figure \ref{fig:pop}.  Fitting the data points of
deuterated methanol, we estimate a mean excitation temperature of 165$\pm$85
K and a mean column density of 9.2$\pm2.0\times10^{16}$ \cms{}.  Since the
data points are scattered due to the low S/N ratio, the fitting results have
a big uncertainty in excitation temperature. Nevertheless, the inferred
excitation temperature is consistent with the temperature at peak abundance
for (deuterated) methanol in warm-up models of hot cores, which is estimated
to be $\sim 120~K$ \citep{Garrod2013} or $\sim 130~K$ \citep{Muller2016}. 
This broad consistency supports the notion that the observed (deuterated)
methanol is produced by warming of icy grains in the disk atmosphere.
If this is indeed the case, the COM-producing icy grains should stay high up
in the disk atmosphere, with interesting implications for grain growth and
settling that should be explored with detailed modeling in future
investigations.

We also measure the integrated line intensity of the methanol lines.  Since
only three lines are detected and the data points are scattered, we can not
estimate the temperature and column density reliably for methanol.  Since
the methanol has the similar spatial distribution to the deuterated
methanol, we assume the same temperature for the methanol, and then estimate
the column density.  Based on the two lines with the upper energy level $E_u
> 300 $ K, which are less optically thick (with a peak brightness
temperature close to 80 K), we can scale down the data points of methanol by
7.4 to roughly match the best-fit solid line of the deuterated
methanol.  This implies a column density of $\gtrsim$
3.4$\pm1.0\times10^{17}$ \cms{} for methanol and thus an abundance ratio
of [\CHtDOH]/[\CHtOH]$\lesssim 0.27$.


Methanol lines have been detected before at $\sim$ \arcsa{0}{55} resolution
toward the center in 19 transitions including the one here at 345.9039 GHz,
with the upper energy level reaching 750 K \citep{Leurini2016}.  The
excitation temperature of the methanol was estimated to be $\sim$ 295 K. 
Thus, the temperature estimated here could be underestimated by a factor of
$\sim$ 2.

The integrated line intensity of \CHtSH{} is $\sim$ 88 K \vkm{} toward the
disk atmosphere in the southern hemisphere.  
Assuming the same excitation temperature of 165 K for
\CHtSH{}, the column density is found to be $\sim 1.0\times10^{17}$ \cms{}, using
the rotation partition function in \citet{Xu2012}.  Thus, 
[\CHtOH]/[\CHtSH]$\sim 3.4$.
The integrated line
intensity is $\sim$ 70 K \vkm{} for \NHtCHO{} and 77 K \vkm{} for \DtCO{}. 
The excitation temperature of \NHtCHO{} and \DtCO{} is unknown.  If we
assume the same excitation temperature, then the column density is $\sim$
$1.6\times10^{15}$ \cms{} for \NHtCHO{} and $\sim$ $3.2\times10^{15}$ \cms{}
for \DtCO{}.















\section{Discussion}

\subsection{Comparing to our previous model}

Our toy model here is adopted from \citet{Sakai2014} and it can be
considered as a modified version of our previous model.  In our previous
model, the infalling envelope was assumed to have a free-fall motion
produced by the gravity of the central source (protostar plus disk) and a
small rotation with conservation of specific angular momentum
\citep{Lee2014}.  Here the infall motion is no longer a pure free-fall, but
a free-fall reduced by the rotational energy according to the conservation
of total energy.  In this case, both the infall and rotation motions produce
constraints on the mass of the central source.  In addition, the infall
velocity reaches a maximum value at the centrifugal radius and then decreases
to zero at the centrifugal barrier, providing a transition region for a disk
to form.  As a result, the disk does not form immediately at the centrifugal
radius as we assumed before, but it can form interior to the centrifugal
barrier.  In our best-fit model, the specific angular momentum is $\sim$ 140
AU \vkm{}, the same as before.  However, the mass of the central source is
$\sim$ 0.25 \solarmass{}, higher than before, which was $\sim$ 
0.18 \solarmass{}.  As discussed earlier, this is
because the mass here gives rise to both the rotation and infall motions,
instead of only the infall motion as assumed in our previous model.  Hence,
the resulting centrifugal radius is $\sim$ 88 AU (\arcsa{0}{22}), smaller
than that estimated before, which was $\sim$ 120 AU (\arcsa{0}{3}).

\subsection{Formation of a Keplerian disk}


From the model results, we can investigate in detail the forming process of
a rotationally supported Keplerian disk in an infalling-rotating
envelope in the star formation.  Figure \ref{fig:model} shows a schematic
diagram of the envelope and disk within $\sim$ 500 AU of the central source
in HH 212.   In our best-fit model to the gas kinematics in \HCOP{} and
COMs, the centrifugal radius (CR) is $\sim$ 88$\pm18$ AU and the centrifugal
barrier (CB) is at $\sim$ 44$\pm$9 AU.  This indicates that \HCOP{} traces
the infalling-rotating envelope extending down to $\sim$ 44 AU and COMs
trace the atmosphere of a Keplerian rotating disk with a radius of $\sim$ 44
AU.  In addition, the innermost envelope in between the centrifugal
radius and the centrifugal barrier can be considered as a transition region,
where the infall velocity decreases rapidly to zero so that the envelope can
transform into a Keplerian disk.  In the schematic diagram, the envelope
structure is derived from the \HCOP{} envelope, as described in Section
\ref{sec:env}.  The disk is composed of two components, one is the dusty
disk seen in continuum in \citet{Lee2017} and the other is the disk
atmosphere (surface) seen here in COMs.  The dusty disk has an outer radius
of $\sim$ 68 AU (\arcsa{0}{17}).  However, based on our model result, only
the inner part within $\sim$ 44 AU can be the Keplerian rotating disk while
the outer part of the dusty disk is in the transition region.

In the dusty disk, the continuum emission is found to become optically thick
at $r \sim$ 40 AU, due to a rapid increase in density \citep{Lee2017}.  
Interestingly, this radius is similar to the radius of the centrifugal
barrier, suggesting that the increase in density is caused by the rapid
decrease of the infall velocity in the transition region,  which is
expected in our model. Since the infall velocity decreases to zero at the
centrifugal barrier, the material there should be almost stagnated, causing
the material to accumulate locally and the disk to become optically thick. 
Also due to this rapid decrease in the infall velocity, the rotation becomes
larger than the infall motion in the transition region, causing the
blueshifted emission and redshifted emission there to be seen on the
opposite sides of the central source, as shown in Figure \ref{fig:HCOP}b. 
In the transition region, the dust continuum emission is only detected in
the midplane, as shown in the schematic model in Figure \ref{fig:model}b,
suggesting that the envelope is denser near the midplane.  This could be due
to a magnetic effect.  If the infalling material is magnetized with an
hour-glass magnetic field morphology, the material would be falling toward
the midplane, as expected for a pseudodisk \citep{Allen2003}.

The dusty disk is flared and in vertical hydrostatic equilibrium
\citep{Lee2017}, as expected for an accretion disk.  More importantly,
its scale height reaches the maximum value of $\sim$ 12 AU at $r \sim$ 36 AU
near the centrifugal barrier, where the peaks of the methanol, deuterated
methanol, and methyl mercaptan emissions are detected.  As discussed
earlier, these emission peaks trace a warm ring in the disk atmosphere with
a temperature of $165-295$ K.  This ring of warm emission could trace a weak
shock, e.g., an accretion shock, produced by the rapid decrease of the
infall velocity near the centrifugal barrier.  Similar weak shocks have been
detected in SO in other protostellar systems L1527 \citep{Sakai2017} and HH
111 \citep{Lee2016}, marking the transition from an infalling envelope to a
Keplerian rotating disk.  The infall velocity there is almost zero (see
Figure \ref{fig:pvCH2DOH}c and \ref{fig:pvCH2DOH}d) and the rotation
velocity interior to that radius can be roughly fitted with a Keplerian
rotation (see Figure \ref{fig:pvmodel}c and \ref{fig:pvmodel}d), indeed
supporting that a Keplerian disk has formed there.

The disk mass integrating from $r \sim$ 36 AU toward the center is $\sim$
0.03 \solarmass{}, using the disk density profile derived from modeling the
disk continuum emission in \citet{Lee2017}, assuming that the continuum emission
is the dust thermal emission.  As discussed by the authors, this mass is an
upper limit because significant continuum emission could come from dust
scattering of the disk thermal emission.  Thus, the mass of the central
protostar is $>$ 0.22 \solarmass{}.  As a result, the disk mass within
$\sim$ 36 AU of the central protostar is $<$ 13\% of the protostellar mass,
further supporting that the Keplerian disk indeed can form with a radius of
$\sim$ 36 AU.

As discussed earlier, the infalling material has to lose a fraction of the
kinetic energy and angular momentum in order to form a Keplerian disk.  In
our observations, the ring of warm molecular gas detected near the
centrifugal barrier supports that a fraction of the kinetic energy has
indeed been converted to thermal energy through the shock, if the ring is
heated by a shock.  In the transition region, a low-velocity outflow or wind
is seen in \HCOP{} extending out from the lower surface in the east (see
Figure \ref{fig:HCOP}b), which may carry away a fraction of the angular
momentum.  This low-velocity outflow may in turn be due to the magnetic
braking \citep{Allen2003}.  Further observations are needed to check if such
an outflow is also seen from other surfaces in the transition region.


\subsection{Comparing to other Class 0 and I disks}



Similar envelope kinematics and disk formation has been suggested in another
Class 0 system L1527 \citep{Sakai2014,Sakai2017}.  In that system, a
geometrically thin infalling envelope is seen in molecular gas traced by
CCH.  It broadens up in the transition region before joining the Keplerian
rotating disk, probably because an accretion shock heats up the material
there \citep{Sakai2017}.  A similar broadening behavior is also seen here in
the transition region in dust continuum, although not in molecular gas
traced by \HCOP{} (see Figure \ref{fig:model}b).  In L1527, the central mass
is $\sim$ 0.18 \solarmass{}, the specific angular momentum in the envelope
is $\sim$ 180 AU \vkm{}, and thus the centrifugal barrier is at $\sim$ 100
AU, about twice as large as in HH 212.  A likely Keplerian rotating disk is
also detected interior to the centrifugal barrier, with a radius of $\sim$
100 AU.  Similar disk radius has also been claimed in other Class 0 systems
\citep{Lee2009,Murillo2013,Yen2017}.

A few Class I Keplerian disks have been claimed in Taurus
\citep{Harsono2014}, with a radius of $\lesssim$ 100 AU, similar to that
found in the Class 0 disks.  According to Equation \ref{eq:CB} in our
model, the centrifugal barrier and thus the disk radius is proportional to
the square of the specific angular momentum in the infalling envelope.
Therefore, the small radius of those disks could be due to the low specific
angular momentum of $\lesssim$ 200 AU \vkm{} in their infalling envelopes,
as in the Class 0 sources.  Recently, a small Class I Keplerian disk
with a slightly larger radius of $\sim$ 100$-$200 AU has also been detected
in HH 111 in Orion \citep{Lee2016}.  However, unlike the Class I sources in
Taurus, HH 111 has a much larger specific angular momentum of $\sim$ 1550 AU
\vkm{} in its envelope.  Its Keplerian disk is not much bigger than those in
the Class 0 sources because a large fraction of the specific angular
momentum is lost, probably due to magnetic braking \citep{Lee2016}. Thus,
the disk radius could be affected significantly by other effects, and thus
becomes much smaller than the centrifugal barrier.

\subsection{Complex Organic Molecules}


Unlike \HCOP{}, the complex organic molecules (COMs) are only detected
interior to the centrifugal barrier within $\sim$ 40 AU of the central
protostar in the atmosphere of the flared disk.  Their emission can be
detected down to $\sim$ 10 AU of the central protostar.  No emission is
detected further in because the disk is bright and optically thick in dust
continuum, and thus the emission is either blocked or absorbed against
it.  The emission of the methanol, deuterated methanol, and methyl mercaptan
peaks near the centrifugal barrier, and thus could be caused by a shock
interaction in that region, as discussed earlier.  On the other hand, since
the disk is flared and thus exposed to the radiation of the central
protostar, stellar radiative heating and/or mechanical heating (by a stellar
or inner-disk wind) could play a role in warming the region where these
molecules are detected.  This is especially true for the molecules detected
at small radii inside the centrifugal barrier, which is less affected by the
envelope-disk transition, if at all.  Formamide and deuterated formaldehyde
are detected closer in and thus their emission is more likely caused by the
radiation heating of the protostar as well.  No emission of these COMs is
detected beyond 40 AU, where the dusty disk is thinning out and thus could
be self-shielded from the radiation (or outflow) of the protostar.









Our observations show the first spatially-resolved detection of these COMs
in gas phase in the disk atmosphere in the Class 0 phase of star formation,
opening up a window to study their formation mechanism.  Methanol is known
to be formed efficiently in the ice mantle of dust grains
\citep{Herbst2009}.  Here the abundance ratio of [\CHtDOH]/[\CHtOH]
($\lesssim 0.27$) is about 4 orders of magnitude higher than the cosmic
abundance of deuterium of D/H $\sim 10^{-5}$.  This high degree of
deuteration strongly supports the formation of methanol and deuterated
methanol in the ice mantle on the dust grains
\citep{Ceccarelli2001,Parise2006}.  We detect these molecules in the gas
phase because they are desorbed (evaporated) from the dust grains due to the
heating by shock and/or the radiation/outflow of the protostar.  This
scenario is supported by the fact that the excitation temperature inferred
from the energy diagram for deuterated methanol is consistent with the
temperature at peak abundance in models for producing this gas phase
molecule from warming of icy grains (see Section 3.4).  Methyl mercaptan has
the same spatial distribution as the methanol and thus could be formed on
the dust grains as well.  Recently, methyl mercaptan has been detected
toward another low-mass star forming region IRAS 16293-2422 and argued to be
formed in the ice mantle on the dust grains and then desorbed
\citep{Majumdar2016}.  Detection of doubly deuterated formaldehyde {\bf
(\DtCO)} is also an indicator of active grain surface chemistry
\citep{Ceccarelli2001}.  Formamide has been detected in both high-mass
star-forming regions \citep{Adande2013} and low-mass star-forming regions
\citep{Kahane2013,Lopez-Sepulcre2015}.  It could be formed through the
interaction between H$_2$CO and NH$_2$ or NH$_4^+$ in gas phase
\citep{Lopez-Sepulcre2015}.  The spatial distribution is similar to that of
\DtCO{} and thus the implied H$_2$CO, supporting this formation mechanism. 
However, this molecule could also be formed on icy grain mantles, e.g., via
hydrogenation of HNCO \citep{Lopez-Sepulcre2015,Mendoza2014}.  It is
detected closer to the protostar than the methanol, which is somewhat
surprising in this scenario because, in the standard warm-up models of hot
cores, its temperature at peak abundance ($\sim 120$ K) is very similar to
that for methanol (and deuterated methanol), which is estimated to be $\sim
120$ K \citep{Garrod2013} or $\sim 130$ K \citep{Muller2016}.  The
temperature at peak abundance for \CHtSH{} is similar to that of NH2CHO
($\sim 120$ K), which is much higher than that of \DtCO{} ($\sim 40$ K; R. 
Garrod, priv.  comm.).




As a result, all of our detected COMs could be formed in the ice mantle on
the dust grains, either in the envelope or in the disk.  Recently, methanol
has been detected in gas phase in a protoplanetary disk (TW Hya) in the
T-Tauri phase of star formation, with its formation in the disk
\citep{Walsh2016}.  Here, our detected COMs are also more likely to be
formed in the disk, either on the surface or in the midplane and then
brought to the surface by turbulence \citep{Furuya2014}.  This is because
the COMs detected here are associated with the warm atmosphere of the disk
rather than a rapidly infalling inner envelope that is warmed up by stellar
radiation or some other means.  This is important because icy grains can in
principle stay longer in the rotationally supported disk than in the
free-falling envelope, which would allow more time for the COMs to form
\cite[e.g.,][]{Jorgensen2005}.  On the other hand, if there is rapid grain
growth, large grains may settle quickly toward the midplane, where the
temperature may be too low to efficiently produce COMs.  Our {\it vertically
resolved} observations provide strong constraints for models of COM
formation and transport.








Methanol has been proposed as a building block for more complex species of
fundamental prebiotic importance, like amino acid compounds
\citep{Walsh2016}.  Formamide has also been proposed as a prebiotic
precursor because it could lead to the synthesis of biomolecules, e.g.,
amino acids and amino sugars, which are the main components for the onset of
both (pre)genetic and (pre)metabolic processes, respectively
\citep{Saladino2012}.  Therefore, the COMs have started to form on disks in
the earliest phase of star formation and may play a crucial role in
producing the rich organic chemistry needed for life.





%



\section{Conclusions}


We have spatially resolved the envelope and disk atmosphere in the HH 212
protostellar system with ALMA.  Our primary conclusions are the following:

\begin{enumerate}

\item The envelope is detected in \HCOP{} down to the dusty disk to $\sim$
40 AU of the central protostar.  It has an infall motion with the infall
velocity increasing toward the center, as expected for a gravitational
collapse.  It also has a rotation motion, which is lower than the infall
motion in large radii but becomes higher than the infall motion near the
dusty disk.

\item A rotating disk atmosphere is detected above and below the dusty disk
in complex organic molecules (COMs), including methanol (\CHtOH{}),
deuterated methanol (\CHtDOH{}), methyl mercaptan (\CHtSH{}), and formamide
(\NHtCHO{}), and in doubly deuterated formaldehyde (\DtCO{}), within $\sim$
40 AU of the central protostar.  These COMs are all spatially resolved,
both radially and vertically, for the first time in the atmosphere of
a disk in the earliest, Class 0 phase of star formation.  These detections
are made possible by ALMA long baseline observations, which resolve both the
dust and molecule line emission in the vertical direction.

\item The kinematics of the envelope and disk atmosphere can be reproduced
simultaneously with a simple toy model, in which the envelope is collapsing
with conservation of both specific angular momentum and total energy and the
disk has a Keplerian rotation interior to the centrifugal barrier.  In the
best-fit model, the envelope has a specific angular momentum of $\sim$ 140$\pm$30
AU \vkm{}, the central source (protostar+disk) has a mass of $\sim$ 0.25$\pm$0.5
\solarmass{}.  The resulting centrifugal radius is $\sim$ 88$\pm$18 AU and the
centrifugal barrier is at a radius of $\sim$ 44$\pm$9 AU.  The innermost envelope
in between these two radii can be considered as a transition region between
a collapsing envelope and a Keplerian rotating disk.  A Keplerian disk can
indeed be formed with a radius of $\sim$ 40 AU.

\item Methanol, deuterated methanol, and methyl mercaptan are detected
mainly in the disk atmosphere at a radius near the centrifugal barrier,
although some are present further in down to a radius of $\sim$ 10 AU. 
Formamide and deuterated formaldehyde are detected within $\sim$ 30 AU of
the central protostar.  

\item Our detection of COMs in the warm disk atmosphere favors their
formation on the disk rather than in the rapidly infall inner (warm)
envelope.  Whether such molecules exist closer to the disk midplane cannot
be determined for the nearly edge-on disk of HH 212 that is optically thick
in dust continuum.  The observed abundances and spatial distributions of the
COMs provide strong constraints on models of their formation and transport.


\end{enumerate}

\acknowledgements

We thank Eric Herbst, Rob Garrod, and {\bf Izaskun Jimenez-Serra} for useful
discussion, {\bf and the anonymous referee for useful comments.}
This paper makes use of the following ALMA data:
ADS/JAO.ALMA\#2012.1.00122.S and 2015.1.00024.S.  ALMA is a partnership of
ESO (representing its member states), NSF (USA) and NINS (Japan), together
with NRC (Canada), NSC and ASIAA (Taiwan), and KASI (Republic of Korea), in
cooperation with the Republic of Chile.  The Joint ALMA Observatory is
operated by ESO, AUI/NRAO and NAOJ.  C.-F.L.  acknowledges grants from the
Ministry of Science and Technology of Taiwan (MoST 104-2119-M-001-015-MY3)
and the Academia Sinica (Career Development Award).  
ZYL is supported in part by NSF1313083 and NASA NNX14AB38G.





\def\nat{Natur}

\begin{figure} [!hbp]
\centering
\putfig{0.7}{270}{f4.eps} 
\figcaption[]
{Redshifted (red contours) and blueshifted (blue contours) emission in the
stacked maps of methanol and deuterated methanol, and in the maps of 
\CHtSH{} (methyl mercaptan), \NHtCHO{} (formamide), and \DtCO{} (deuterated
formaldehyde), on top of the dusty disk continuum map.  The maps all
have an angular resolution of $\sim$ \arcsa{0}{04}.
The upper energy
level for the line transition in each molecule
is indicated in the upper right corner.  The contour levels all start
at 3$\sigma$, with a step of 2$\sigma$ for methanol and deuterated methanol
and a step of 1 $\sigma$ for the rest.  
The noise levels $\sigma$ are $\sim$ 2.8 \mJybk{} for methanol,  
1.0 \mJybk{} for deuterated methanol,
3.0 \mJybk{} for methyl mercaptan,
3.5 \mJybk{} for formamide,
and 4.3 \mJybk{} for deuterated
formaldehyde.
\label{fig:complexmol}} 
\end{figure}

\begin{figure} [!hbp]
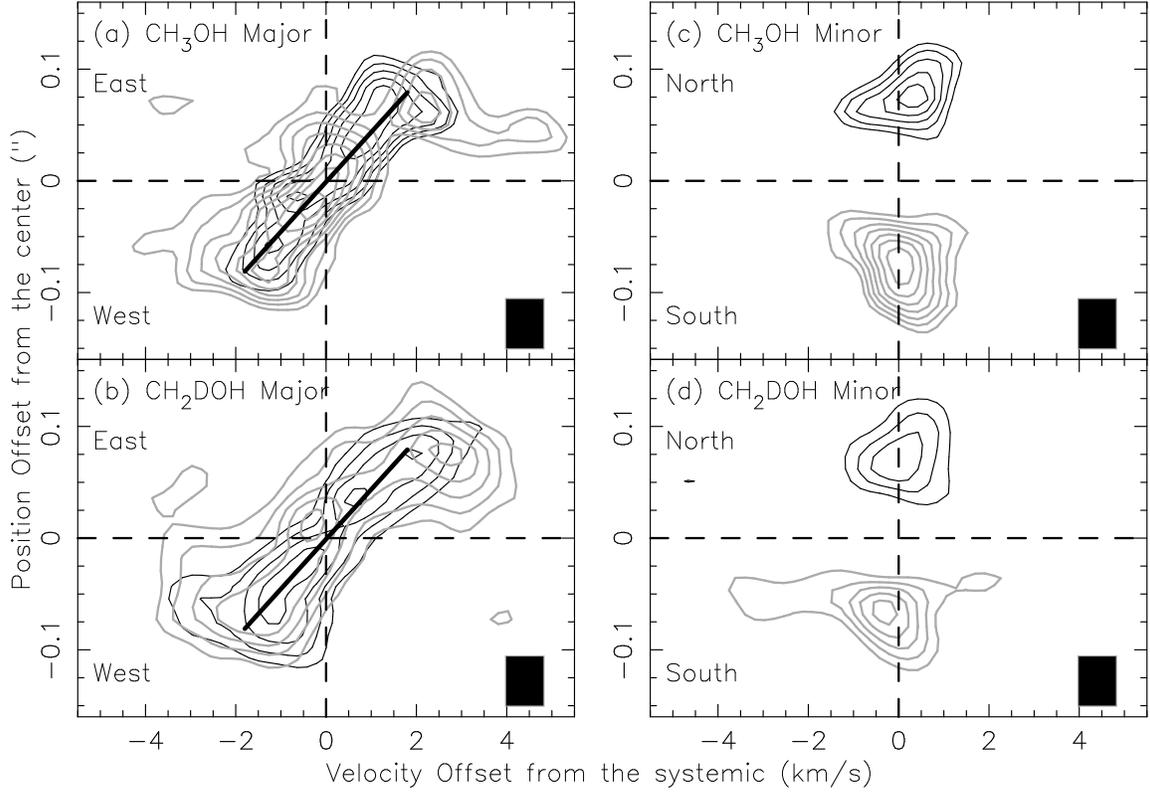

\centering
\putfig{0.6}{270}{f5.eps} 
\figcaption[]
{PV diagrams of the disk atmospheres in methanol and deuterated methanol.
Black contours are for the upper disk atmosphere and
gray contours are for the lower disk atmosphere.
The horizontal dashed line indicates the position of the
protostar.  The vertical dashed line indicates the systemic velocity.
(a) and (b) are the PV diagrams for cuts along the major axis.
The solid lines show the linear fit to the emission peaks
in the linear PV structures seen in both methanol and deuterated methanol.
(c) and (d) are the PV diagrams for cuts along the minor axis.
The contour levels start at 3$\sigma$, with a step of 1$\sigma$ for methanol
and a step of 2$\sigma$ for deuterated methanol. The noise levels
$\sigma$ are 0.93 \mJyb{} for methanol and 0.35 \mJyb{} for deuterated methanol.
\label{fig:pvCH2DOH}}
\end{figure}

\begin{figure} [!hbp]
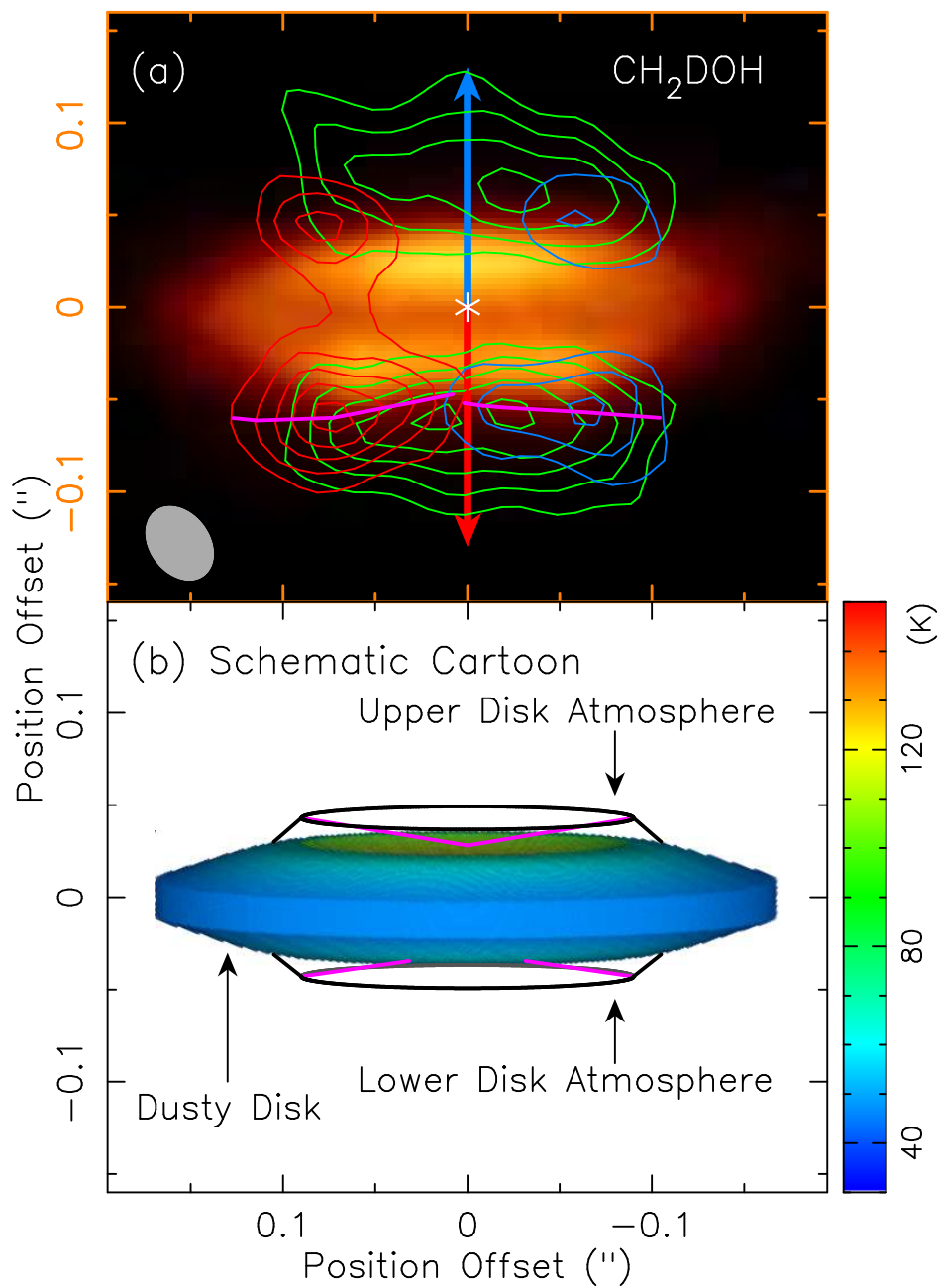

\centering
\putfig{1}{270}{f6.eps} 
\figcaption[]
{(a) The deuterated methanol maps for the linear PV structure (green contours) and
the differential PV structure (red for redshifted emission and blue for blueshifted emission).
The contour levels start at 3$\sigma$, with a step of 3$\sigma$, where
$\sigma=0.56$ \mJyb{}. The magenta lines show the slight
decrease of scale height toward the center for the redshifted and blueshifted
emission that trace the differential PV structure.
(b) A cartoon showing a disk atmosphere of complex organic molecular gas on
top of the dusty disk surface.  The dusty disk model is adopted from
\citet{Lee2017}, showing the scale height at different radius, with the
color indicating the surface temperature of the dusty disk.
The disk atmosphere is assumed to be flared, with
the magenta lines showing its scale height along the
major axis.
\label{fig:com}}
\end{figure}

\begin{figure} [!hbp]
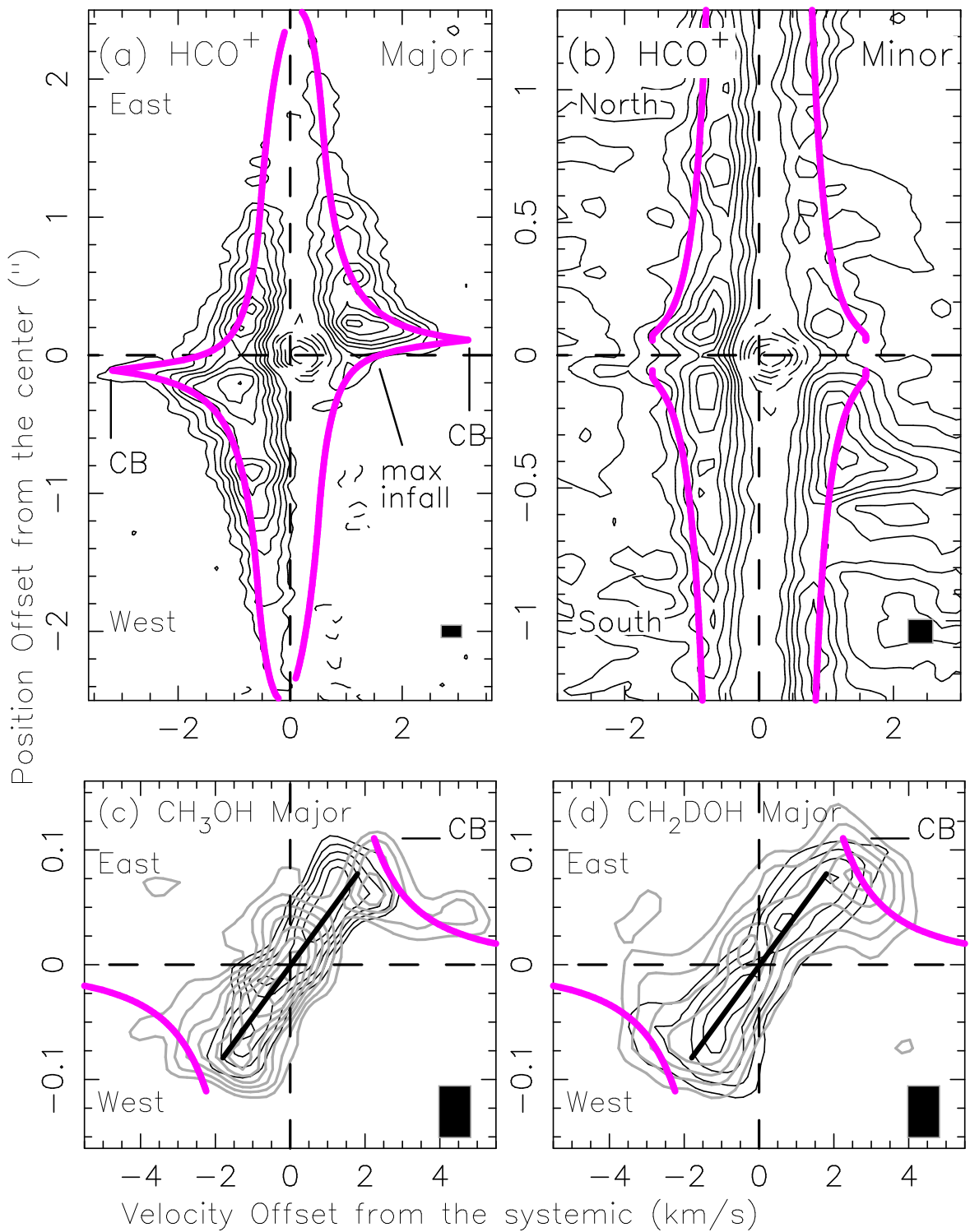

\centering
\putfig{0.85}{0}{f7.eps} 
\figcaption[]
{Fitting (purple curves) to the PV diagrams of the \HCOP{} envelope cut
along the major and minor axis, and the PV diagrams of \CHtOH{} and
\CHtDOH{} disk atmosphere along the major axis.
The contour levels are the same as in Figures \ref{fig:pvHCOP} and \ref{fig:pvCH2DOH}.
CB means centrifugal barrier.
\label{fig:pvmodel}}
\end{figure}

\begin{figure} [!hbp]
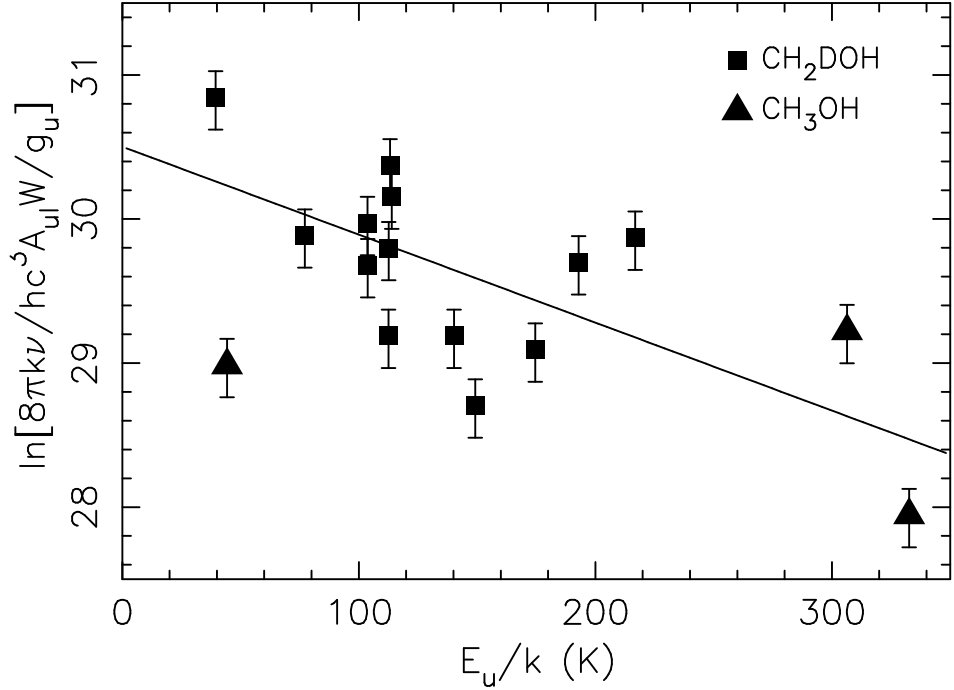

\centering
\putfig{1}{270}{f8.eps} 
\figcaption[]
{The population diagram for \CHtDOH{} (squares) derived from the total intensity map
for the lower disk atmosphere in the south. 
The error bars show the uncertainty in our measurements, which are assumed 
to be 20\% of the data values.
The solid line is a linear fit to the data. 
Moreover, 3 data points of \CHtOH{} (triangles) are also included to estimate its column density. 
The data points of methanol have been scaled down (i.e., divided) by 7.4
to roughly match the best-fit solid line of the deuterated methanol.
\label{fig:pop}}
\end{figure}

\begin{figure} [!hbp]
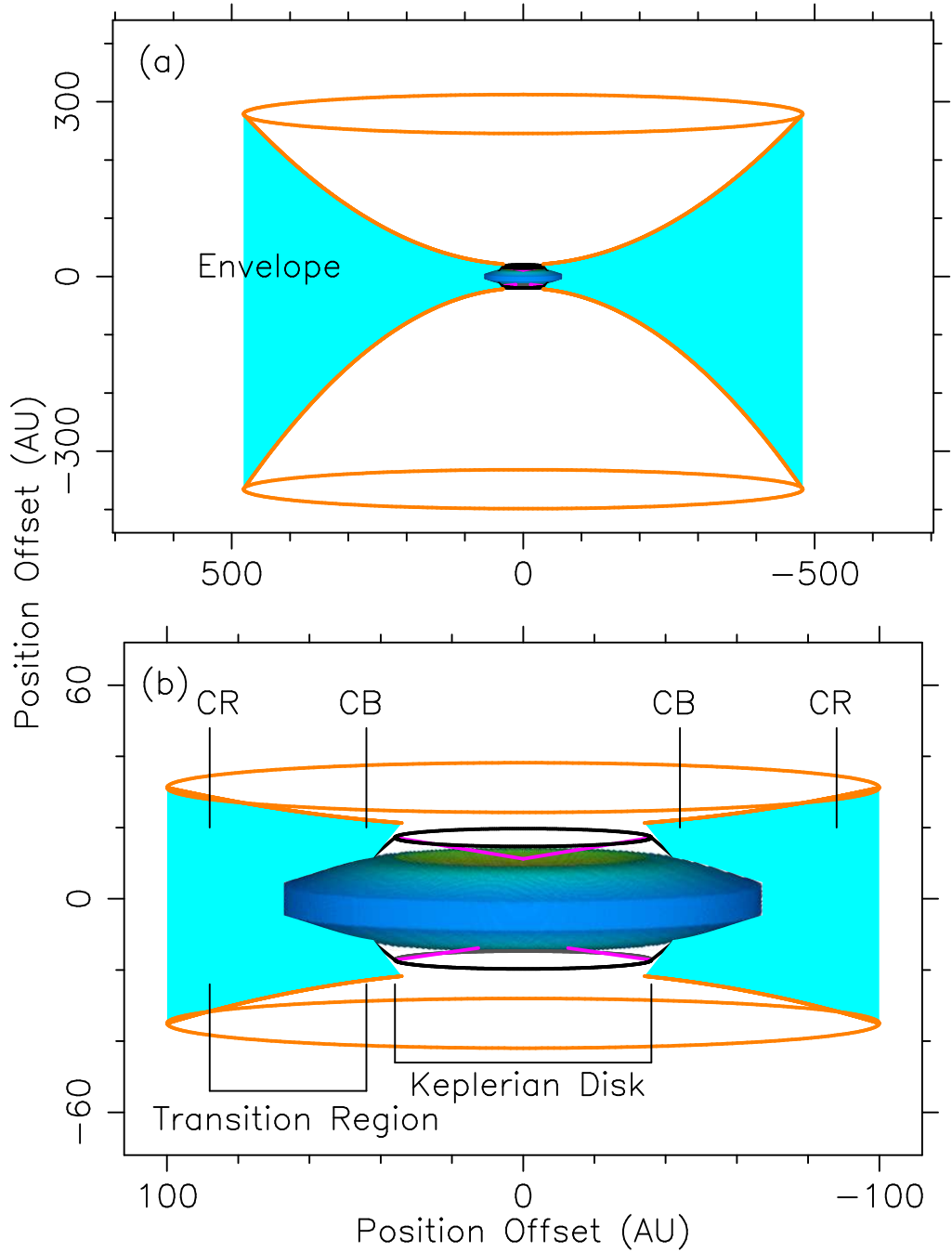

\centering
\putfig{1}{270}{f9.eps} 
\figcaption[]
{Schematic model for the envelope and disk in HH 212. 
Within $\sim$ 36 AU of the center, the dusty disk is flared \citep{Lee2017} and
the disk atmosphere rotates with a Keplerian rotation, indicating that a Keplerian
is formed there. The transition region is the region in between
the centrifugal barrier (CB) and the centrifugal radius (CR).
\label{fig:model}}
\end{figure}

\clearpage

\begin{table}
\small
\centering
\caption{Observation Logs}
\label{tab:obs}
\begin{tabular}{llllll}
\hline
Cycle & Date & Array &Number of &Time on target & Baselines  \\
      & (YYYY-MM-DD) & Configuration & Antennas&(minutes) & (meter) \\
\hline\hline
1 & 2015-08-29 & C32-6   & 37 & 30  & 15$-$1466   \\ 
1 & 2015-08-29 & C32-6   & 35 & 30  & 15$-$1466   \\ 
3 & 2015-11-05 & C36-7/8 & 45 & 44  & 78$-$16196 \\ 
3 & 2015-12-03 & C36-7/8 & 32 & 44  & 17$-$6344  \\ 
\hline
\end{tabular}
\end{table}

\begin{table}
\small
\centering
\caption{Correlator Setup for Cycle 1 Project}
\label{tab:corr1}
\begin{tabular}{llllll}
\hline
Spectral  & Line or   & Number of & Central Frequency & Bandwidth & Channel Width$^a$\\
Window & Continuum & Channels  & (GHz)             & (MHz)     & (kHz) \\
\hline\hline
0 & CO J=3-2      & 3840  & 345.803 & 468.750  & 122.070  \\
1 & SiO J=8-7     & 3840  & 347.338 & 468.750  & 122.070  \\
2 & HCO$^+$ J=4-3 & 3840  & 356.742 & 468.750  & 122.070  \\
3 & Continuum     & 3840  & 358.008 &1875.000  & 488.281  \\
\hline
\multicolumn{6}{l}{a: Effective Velocity Resolution is about 2 Channels in this Cycle}
\end{tabular}
\end{table}

\begin{table}
\small
\centering
\caption{Correlator Setup for Cycle 3 Project}
\label{tab:corr3}
\begin{tabular}{llllll}
\hline
Spectral  & Line or   & Number of & Central Frequency & Bandwidth & Channel Width\\
Window & Continuum & Channels  & (GHz)             & (MHz)     & (kHz) \\
\hline\hline
0 & SO \SOt      & 960   & 346.528 & 234.375  & 244.140  \\
1 & CO $J=3-2$      & 960   & 345.796 & 234.375  & 244.140  \\
2 & H$^{13}$CO$^+$ $J=4-3$      & 960   & 346.998 & 234.375  & 244.140  \\
3 & SiO $J=8-7$     & 960   & 347.330 & 234.375  & 244.140  \\
4 & HCO$^+$ $J=4-3$ & 1920  & 356.735 & 468.750  & 244.140  \\
5 & Continuum     & 1920  & 357.994 &1875.000  & 976.562  \\
\hline
\end{tabular}
\end{table}

\begin{table}
\small
\centering
\caption{Calibrators and Their Flux Densities}
\label{tab:calib}
\begin{tabular}{llll}
\hline
Date & Bandpass Calibrator &Flux  Calibrator & Phase Calibrator \\
(YYYY-MM-DD) & (Quasar, Flux Density) & (Quasar, Flux Density) & (Quasar, Flux Density) \\
\hline\hline
2015-08-29 & J0607-0834, 1.20 Jy & J0423-013, 1.03 Jy & J0552+0313, 0.25 Jy  \\ 
2015-08-29 & J0607-0834, 1.20 Jy & J0423-013, 1.03 Jy &  J0552+0313, 0.25 Jy \\ 
2015-11-05 & J0423-0120, 0.55 Jy & J0423-0120, 0.55 Jy & J0541-0211, 0.22 Jy \\  
2015-12-03 & J0510+1800, 4.07 Jy & J0423-0120, 0.67 Jy & J0541-0211, 0.23 Jy \\ 
\hline
\end{tabular}
\end{table}

%
\begin{deluxetable}{lcccccc}
\tablecolumns{8}
\tabletypesize{\normalsize}
\tablecaption{Line Properties from Splatalogue
 \label{tab:lines}}
\tablewidth{0pt}
\tablehead{
\colhead{Molecule} & \colhead{Frequency} & Transition & 
$S_{ij}\mu^2$ &     log$_{10}$($A_{ij}$) &     $E_{up}$ & Linelist \\
               & \colhead{(GHz)}   & QNs  & (D$^2$)            &   (s$^{-1}$) & (K) & 
}  
\startdata
 \CHtDOH & 345.71872 & 3(2,1) - 2(1,2), e1 &	0.61635 & 	-4.37316 & 	39.43488  & 	JPL \\
 \CHtDOH & 356.89967 & 8(2,7) - 7(2,6), e1 &	5.77817 &	-3.74509 &	103.67562 &	JPL \\
 \CHtDOH & 356.90507 & 8(5,3) - 7(5,2), o1 &	3.71100 &	-3.93737 &	192.93386 &	JPL \\
 \CHtDOH & 356.90507 & 8(5,4) - 7(5,3), o1 &	3.71100 &	-3.93737 &	192.93386 &	JPL \\
 \CHtDOH & 356.91471 & 8(2,7) - 7(2,6), o1 &	5.99850 &	-3.72878 &	112.57969 &	JPL \\
 \CHtDOH & 356.93242 & 8(4,4) - 7(4,3), e1 &	4.51361 &	-3.85223 &	149.20816 &	JPL \\
 \CHtDOH & 356.93244 & 8(4,5) - 7(4,4), e1 &	4.51361 &	-3.85223 &	149.20816 &	JPL \\
 \CHtDOH & 357.23327 & 8(2,6) - 7(2,5), o1 &	6.01053 &	-3.72675 &	112.61829 &	JPL \\
 \CHtDOH & 357.52856 & 7(1,6) - 7(0,7), e1 &	6.14490 &	-3.66171 &	77.13300 &	JPL \\
 \CHtDOH & 357.53536 & 8(6,2) - 7(6,1), e0 &	2.19644 &	-4.16284 & 	216.82322 & 	JPL \\
 \CHtDOH & 357.53536 & 8(6,3) - 7(6,2), e0 &	2.19644 &	-4.16284 &	216.82322 &	JPL \\
 \CHtDOH & 357.65950 & 8(5,4) - 7(5,3), e0 &	2.96147 &	-4.03260 & 	174.61159 &	JPL \\
 \CHtDOH & 357.65950 & 8(5,3) - 7(5,2), e0 &	2.96147 &	-4.03260 &	174.61159 &	JPL \\
 \CHtDOH & 357.68738 & 8(2,6) - 7(2,5), e1 &	5.77277 &	-3.74262 &	103.77011 &	JPL \\
 \CHtDOH & 357.81966 & 8(4,5) - 7(4,4), e0 &	3.57036 &	-3.95081 &	140.21085 &	JPL \\
 \CHtDOH & 357.82025 & 8(4,4) - 7(4,3), e0 &	3.57035 &	-3.95081 &	140.21087 &	JPL \\
 \CHtDOH & 357.98957 & 8(3,6) - 7(3,5), e0 &	3.94436 &	-3.90693 &	113.18953 &	JPL \\
 \CHtDOH & 358.05489 & 8(3,5) - 7(3,4), e0 &	3.94297 &	-3.90684 &	113.19525 &	JPL \\
 \CHtOH  & 345.90392 & 16(1,15) - 15(2,14) & 	6.18282 & 	-4.04453 & 	332.65331 &	JPL \\
 \CHtOH  & 358.41465 & 10(6,5) - 11(5,7) 	& 1.34748 	&       -4.46361 &	306.42756 &	JPL \\
 \CHtOH  & 358.60580 & 4(1,3) - 3(0,3) 	& 2.21223 	&       -3.87963 &	44.26398 &	JPL \\
 \CHtSH  & 356.62709 & 14(1) - 13(1) E &	0.24000 & 	-4.00147 &	135.98637 &	SLAIM \\
 \NHtCHO & 356.71376 & 17(2,16) - 16(2,15) & 218.95428 	&       -2.48082 & 	166.78985 & 	CDMS \\
 \HCOP   & 356.73422 & 4 - 3 &	60.83879 & -2.44709 	& 42.80203 &	CDMS \\
 \DtCO   & 357.87145 & 6(2,4) - 5(2,3) &	57.94797 &	-2.92482 &	81.21173 &	CDMS\\
\enddata
\end{deluxetable}
\clearpage

\end{document}